# JSAnalyzer: A Web Developer Tool for Simplifying Mobile Pages Through JavaScript Optimizations

MOUMENA CHAQFEH, New York University Abu Dhabi, UAE
JACINTA HU, New York University Abu Dhabi, UAE
WALEED HASHMI, New York University Abu Dhabi, UAE
RUSSELL COKE, New York University Abu Dhabi, UAE
LAKSHMI SUBRAMANIAN, New York University, USA
YASIR ZAKI, New York University Abu Dhabi, UAE

The amount of JavaScript embedded in Web pages has substantially grown in the past decade, leading to large and complex pages that are computationally intensive for mobile devices. In this paper, we propose *JSAnalyzer*, an easy-to-use tool that enables Web developers to quickly optimize and generate simpler versions of existing web pages for mobile users. *JSAnalyzer* can selectively enable or disable JavaScript elements in a page while visually observing their impact, such that non-critical elements can be removed without sacrificing the visual content or the interactive functionality. Our quantitative evaluation results show that *JSAnalyzer* achieves more than 88% relative increase in performance scoring for low-end mobile phones (i.e., from 32% to 60%), and reduces the page load time by 30%. A qualitative study of 22 users shows that *JSAnalyzer* maintains more than 90% visual similarity to the original pages, whereas a developer evaluation study conducted with 23 developers shows that *JSAnalyzer* scores more than 80% in terms of usefulness and usability while retaining the page content and functional features. Additionally, we show that *JSAnalyzer* outperforms state-of-the-art solutions such as JSCleaner and Google AMP.

## 1 INTRODUCTION

Mobile web user experience has become a major concern due to two main reasons: the increasing complexity of the pages, and the processing limitations of mobile devices. The average Page Load Time (PLT) of existing mobile pages is about 19 seconds on a 3G connection and 14 seconds on a 4G connection, while 53% of a mobile page visitors will leave the page if it doesn't load within three seconds [24]. These numbers suggest that many web developers might be unaware of the performance of their pages on mobile devices.

The processing cost of pages is even higher when accessed by low-end mobile devices (with low RAM and CPU), which are popular in developing regions. The total number of mobile web users has recently exceeded 3.5 billion worldwide with 74% of users living in low and middle income countries [15], where smartphones are the primary (and sometimes the sole) means of Internet access [8]. Many of these countries are plagued with poor network infrastructure, where users suffer from low data rates and extremely long network delays.

While the provision of fast and affordable connectivity worldwide would be the ideal solution to bridge the access and performance gap between the under-served users and the developed world, the current plans in many countries (see example [11]) shows that applying this solution would take years to be achieved. A faster and cheaper alternative is to provide lighter versions of web pages that are less intensive to be processed by mobile browsers.

When it comes to browser processing, JavaScript (JS) elements used by most modern web pages [30] are proven to be the most expensive compared to other equivalent sized resources [45]. Despite its impact on web pages' performance, the current status of the World Wide Web shows that mobile pages utilize 20 external JS elements at the $50^{th}$ percentile of the number of JS distribution taken over seven million web pages (in comparison to 21 external JS elements for Desktop pages), where a triple processing time can be witnessed on a mobile device compared to a desktop [12]. While best practices show that considering smaller JS bundles is the ideal way to improve the performance of a web page [12], existing developer tools do not consider JS optimizations, but rather focus on other considerations such as: tracking and debugging JS code [29, 34, 40], borrowing functionality from existing pages [49], or reducing the effort to reproduce contents [27].

In this paper, we propose *JSAnalyzer* to provide a tool for Web developers to rapidly prepare lightweight versions of existing pages, with the objective of improving mobile web access for hundreds of millions of users in developing regions. Since JS may (or may not) be utilized in web pages for visual content generation, and/or interactive functionality provisioning (such as interactive menus), a major challenge that is addressed by *JSAnalyzer* is the ability to maintain the visual content and functionality of the pages. With *JSAnalyzer*, each JS element on a web page can be enabled or disabled via an interactive user interface. The effect of the user interactions (enabling/disabling) on the page is visualized in a controlled web browser, which loads a new version of the page according to these interactions upon request—these reloads happen almost instantaneously given the fact that there are being requested from a close-by cache. This aids the developer in making a final decision on JS preservation or removal based on the visual impact observed on the page. As such, all JS elements that do not sacrifice the page content or functionality can be disabled.

To aid the user in identifying the functionality of JS elements (especially when they have no visual effect on the page), we integrate a JS classifier that attempts to tag each element with an appropriate category. The considered categories are suggested by the experts of the web community [16], which are: Advertising, Analytics, Social, Video, Utilities, Hosting, Marketing, Customer Success, Content, CDN, and Tag Management. These tags aim to assist the developer in making a decision on what to preserve and what to remove. These tags are merely used to give the developer a suggestion about the category behind a certain JS element, however the final removal decision is left to the developer.

By the end of an analysis session, the developer can save and evaluate a simplified version of the page, which preserves all JS elements enabled by the developer, and removes disabled elements. The simplified page is characterized by an improved performance in comparison to the original page, not only in terms of PLT, but also in bandwidth and energy consumption due to the reduction in the download size. With a reduced bandwidth consumption, users can browse more pages using their data plans, while the reduction in consuming energy can increase the life time of mobile battery (since web browsing is a major drain on the battery power) [42].

We evaluated 100 popular pages simplified by *JSAnalyzer* quantitatively (using a low-end and a high-end mobile phone) and qualitatively (via a qualitative evaluation tool and a user study). In addition, we evaluated *JSAnalyzer* by recruiting 23 developers with various web development experience. The quantitative evaluation shows an improvement in the overall user experience metrics on low-end phones by more than 88% (from 32% to 60%), and by 38% (from 45% to 62%) on high-end phones. Results also show a reduction in the page load time of around 30%. The qualitative user study of 22 users shows that the simplified pages maintain more than 90% visual similarity to the original pages.

*JSAnalyzer* provides better results in terms of the user experience metrics in comparison to *JSCleaner* [26] and *Google AMP* [32]. In contrast to JSCleaner which removes JS elements from web pages based on their features, *JSAnalyzer* provides an interactive interface via which the developer can qualitatively evaluate the impact of JS removal on the page. With JSAnalyzer, the simplification process happens one step at a time, so that the developer can disable a certain JS element and immediately observe the visual impact of this action. On the other hand, JSCleaner blindly removes non-critical elements all at once, and the user can only observe the final simplified version. Consequently, the pages simplified using JSAnalyzer show higher similarity scores to the original pages compared to JSCleaner pages. In summary, the contribution of this paper includes:

- *JSAnalyzer*, a novel JS analysis environment that works on all web pages to rapidly prepare lighter versions for mobile web, without sacrificing the visual content or the functionality of the original pages.
- An evaluation of *JSAnalyzer* pages in terms of performance gains, the overall user experience, and the pages quality.
- An evaluation study of *JSAnalyzer* tool from different perspectives depicted by 23 web developers.

## 2 RELATED WORK

### 2.1 JavaScript Tools for Web Developers and Users

To the best of our knowledge, current developer tools do not allow for the dynamic generation of lighter versions of web pages from existing original pages, which became a crucial requirement for developers to accommodate the requirements of mobile web users. *JSAnalyzer* aims to fill the gap in existing developer tools [23, 27, 29, 34, 35, 40, 49] that allow for JS debugging and tracking in web pages, or producing interactive content from existing pages. Chrome Developer Tools are extended in [34] to track and visualize JS method calls and JS libraries, where JS libraries are detected in existing pages and listed to the user with their corresponding versions. In [23] and [39], it is proposed to assist developers in identifying and locating the code that implements interactive features in web pages, while in [35] and [44], a better understanding is provided to discover how JavaScript supports interactivity in web pages.

JS is also utilized in developer tools for easing content generation [27, 49]. With [49], developers can borrow functionality from existing web pages, by extracting components from these pages and turning them into self-contained widgets that can then be embedded into other pages. In [27], a simple markup language is combined with reactive JS components to reduce the effort needed to produce interactive documents.

In trying to handle the cost of JS, web developers utilize uglifiers [2, 21] to reduce the size of JS files before embedding them into their pages. These uglifiers (also known as minifiers) remove all unnecessary characters (including white spaces, new lines and comments) from JS files without changing functionality. Despite the potential enhancement in JS transmission efficiency, the page performance will remain sacrificed because the browser has to interpret the entire JS. In contrast, *JSAnalyzer* optimizes JS usage in web pages by removing elements that are non-critical to the page content or interactive functionality. On the other hand, to help web users in reducing the amount of JS transferred to their browsers, web developers provide in-browser blocking extensions [5, 46] that either eliminate the whole JS or prohibit a specific category of JS such as Ads or trackers [1, 20, 36]. Some of these blockers rely on users to configure what to block. Despite that the Ad and tracking blockers can rely on automatic updates of blocking lists, they are still unable to predict if an unknown JS falls into the target category or not.

### 2.2 Web Complexity Solutions

Part of today's problem in web development is the fact that it's mainly considered for the developed world, where high-end phones are capable of processing all of the unnecessary JS code with a negligible impact on the overall performance or the user experience. This unfortunately does not hold for a large portion of users in developing regions with low-end devices. It's worth here to mention the fact that although processing these unnecessary JS code might not affect the user experience in the developed world, it affects the mobile phone power/battery consumption, which is an issue that many users suffer from. Surprisingly, many developers do not evaluate their pages using low-end devices. According to our developer's study (see Section 5.4), a high percentage of developers are aware of the necessity to provide lighter mobile versions of web pages, but only 30% of them have actually evaluated their pages on mobile devices.

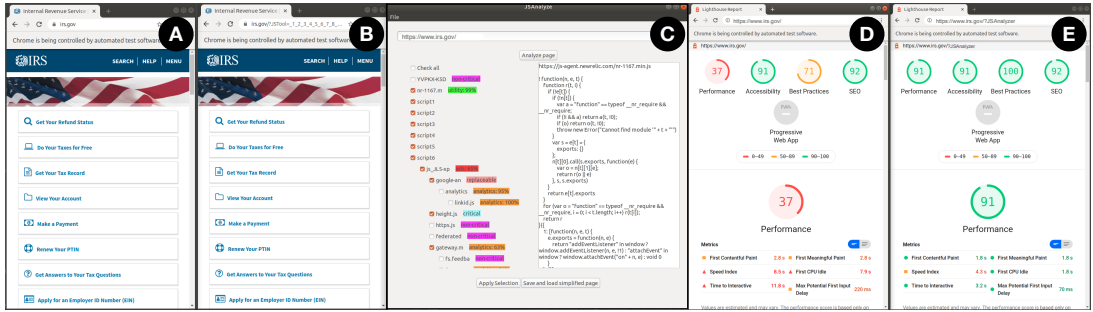

Fig. 1. JSAnalyzer Session [A: Original page, B: Simplified page, C: Interactive interface, D: Original page performance report, E: Simplified page performance report.]

During the last decade, an increasing interest is shown to tackle the complexity of web pages [43], including developer tools, in-browser tools, platform-based solutions, and new browsers. For instance, Ply [38] is an example tool that helped developers in speeding up replicating complex web features by 50% compared to prior tools. On the other hand, SpeedReader [31] is an in-browser tool that aimed at enhancing pages that are suitable for the reader mode of web browsers. However, it did not consider pages that utilize JS for content generation. Wprof [47] is another in-browser tool that provided an understanding of the key hindering effects behind the PLT. It showed that JS has a considerable impact on the PLT due to its role in blocking HTML parsing. To speedup PLT, Shandian [48] restructures the page loading process, whereas Polaris [41] detects additional edges to allow for more accurate fetch schedules.

In addition to PLT improvement, different platform-based solutions have been proposed to improve the overall user browsing experience. For example, Apple News [3] is provided for Apple mobile devices, and Instant Articles [4] is provided via the Facebook applications. These solutions only benefit the users of these specific applications. Other approaches and products have involved the utilization of a proxy server to aid web speed boosting for an improved performance, such as Amazon Silk [7] and Opera Mini [22].

Accelerated Mobile Pages (AMP) is the Google approach to tackle the complexity of modern web pages (AMP) [32], which is recently characterized in [37]. A major difference between JSAnalyzer and AMP is that the latter provides a framework for developers to create new pages, whereas JSAnalyzer aims to rapidly optimize existing pages for mobile web. A recent solution named *JSCleaner* [26] generates simplified web pages and provides them via a proxy server. However, it does not provide a methodology to guarantee the content completeness or functionality preservation in the simplified pages. In contrast, *JSAnalyzer* allows to visually evaluate the quality of the page to generate with respect to the corresponding original version.

## 3 JSANALYZER DESIGN

*JSAnalyzer* is a tool that benefits web developers in two ways: First, it gives an insight into the functionality of each JS elements embedded in a web page. This is achieved through an initial category prediction provided by the tool (see Section 3.2), and an analysis phase where the user can interact with these elements by enabling and disabling (see Section 3.4). The objective of the user interactions is to evaluate the impact of each of these elements on the page, and check which elements are critical to the page and which elements are non-critical. Second, it generates a lighter version of the analyzed page, by removing the non-critical elements. *JSAnalyzer* enables the developer to evaluate the quality of the newly generated pages before making them available for mobile access. This is done by comparing the reports generated by the Google lighthouse tool [18]. Figure 1 shows an analysis session of *JSAnalyzer*.

JSAnalyzer system consists of an interactive User Interface (UI) and a proxy server with a database. The UI integrates the JS classifier 3.2 in addition to two evaluation tools 3.4 (mobile page generation and evaluation). The proxy server performs JS optimization and generates new temporal versions of the page during the analysis sessions (according to the user interactivity), and then caches the *JSAnalyzer Page* upon request. Figure 2 shows the main components in *JSAnalyzer* architecture. *JSAnalyzer* is inspired by three factors: the cost of JS in existing web pages, the need for fast mobile access to these pages, and the lack of tools to understand what the different JS elements embedded in web pages are supposed to provide.

A JS element refers to any JavaScript code used by the page. There are three different kinds of these elements: 1) inline JS code within the main HTML index file written within a <script> tag (without a src attribute being defined), 2) external JS code defined within the main HTML index file by specifying a src URL within a <script> tag pointing to an external .js file, 3) a recursive JS code, being brought by another external JS code element. The src URL is usually defined within the parent JS file. The latter kind is usually the most difficult kind to track given that there is no indication on the existence of this JS script within the main HTML index. The browser needs to evaluate and process the page in order to figure out these recursive JS. A large number of today's JS libraries tend to bring multiple recursive JS elements that the original web developer is unaware of, thus increasing the pages complexity.

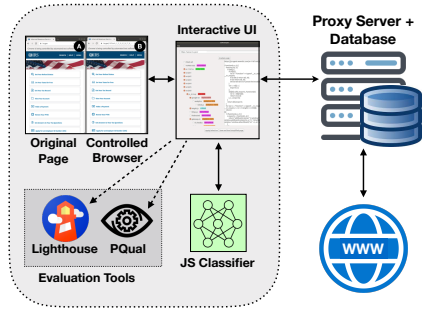

Fig. 2. JSAnalyzer Architecture

Without *JSAnalyzer*, web developers who intend to provide mobile web users with faster versions of their pages should create new pages from scratch. However, *JSAnalyzer* allows for the rapid generation of lightweight content from existing pages. This is achieved by reducing the number of JS elements embedded in the original pages, without sacrificing the visual content or interactive functionality. The design of *JSAnalyzer* makes it very easy for any developer, even novice, to use the tool and be able to visually decide on the criticality of the JS elements. Thus, optimizing and improving the overall pages' efficiency without having to deal with understanding the complex JS elements' code.

### 3.1 Design Considerations

Today's common web development practices hinder the task of JS optimization in modern web pages, since reverse engineering is an extremely complex and time-consuming task [34]. Web developers often rely on existing frameworks which tend to add a number of JS libraries by default. As a result, many developers are unaware of what's happening behind the scene. *JSAnalyzer* gives the control back to developers to investigate JS elements embedded in their pages, and to help them optimize these pages for mobile access. *JSAnalyzer* does not assume special development skills. It works for all web pages regardless of their complexity, and eases the process of generating mobile friendly web pages.

*3.1.1 JavaScript is not always essential.* Identifying which JS elements are non-critical to the overall page content is a challenging task. Many web developers today rely heavily on ready made frameworks, tools, and external libraries for implementing certain features in their sites. Most of these come in the form of a black box approach, in the sense that the developer will import a certain library and use a set of functions to add certain functionalities to a given web page. What most of the developers are unaware of, or are deliberately ignoring is the fact that a lot of these libraries end up recursively bringing additional JS files that are not directly requested by the index.html of the main page. This is one of the main issues that is addressed by *JSAnalyzer*, where the developer can analyze and visually evaluate the full list of all JS elements brought to a given web page (including the ones that are being recursively brought by the browser, and are not part of the main page source-code—making it difficult for developers to spot or comment out). To investigate the user preferences in removing non-critical JS elements, a user survey of mobile users in developing regions is conducted in [25], where users are asked to select all non-essential elements which they are willing to remove. Survey results show that 99.6% of the participants are willing to remove some JS elements from web pages to browse lighter versions of these pages (i.e., they identified at least one category as non-essential in responding to the related survey question). It also shows that most of the users are willing to remove advertising and analytics JS. The default configurations of *JSAnalyzer* is inspired by these results to consider both advertising and analytics as non-critical elements that should be blocked.

*3.1.2 The Dynamic Nature of JavaScript.* The dynamic nature of JS programming makes it extremely difficult, even for well versed developers, to keep track of these files and validate if they are actually required for the expected functionality. The problem with these libraries is that they are generic sets of functions that are meant to provide different functionalities. These libraries come as a bundle of complex sets of large JS files (in today's web the heavy use of NodeJS adds to this bundle complexity issues, due to the automatic generation of these bundles). Unlike conventional programming, where a compiler removes the unused functions, JavaScript is dynamically interpreted in the browsers, and static analysis of unused code is not possible. The alternative approach provided by *JSAnalyzer* is to visually evaluate the impact of JS elements by enabling/disabling them, in order to assist developers in making decisions on which elements can be safely removed without affecting the page content and functionality from the end-user perspective.

*3.1.3 Handling Recursive JavaScript.* Looking at the library bundles, It's often the case that many of these functions are not used by the page (but are being processed by the phone browser), and many of the recursive JS files might not even be required for the targeted set of implemented page functionality. So *JSAnalyzer* would shed a light on these external recursive JS files, on whether they are really required by the page or not.

*3.1.4 Retaining The Pages' Content and Functionality.* JSAnalyzer aims to strike a certain balance between retaining as many of the original content of the pages as possible, and speeding up the page load time (i.e., enhancing the overall user experience). This means that certain sacrifices/compromises need to be made. Explaining this balance to our human evaluators in our web developers evaluation, the majority of them decided to sacrifice JS elements that are not directly affecting the main page content, such as tracking and advertising.

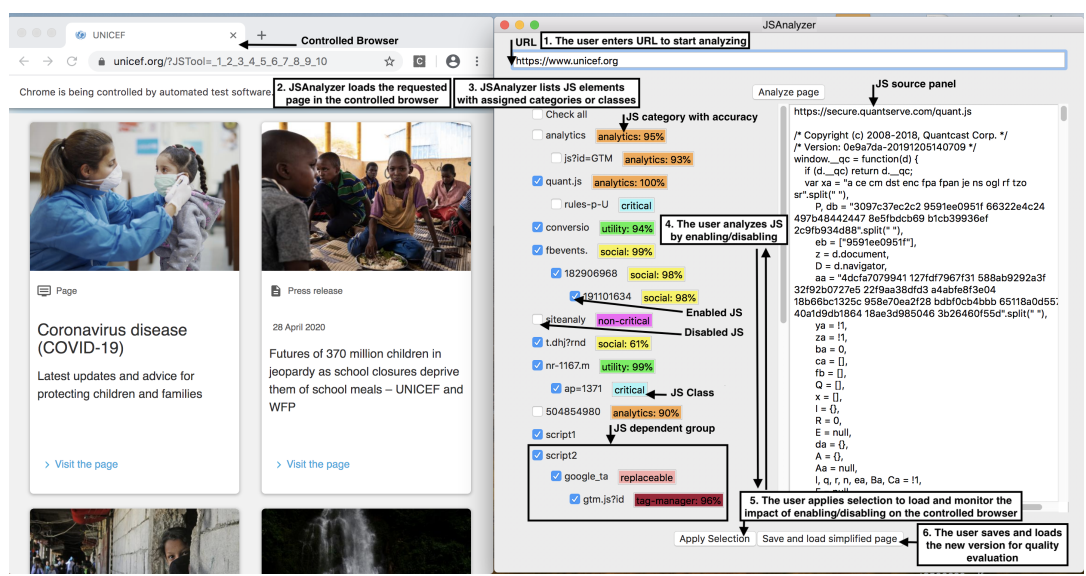

Fig. 3. JSAnalyzer User Interface: From entering a URL to generating a lightweight page for mobile web.

## 3.2 JavaScript Classifier

*JSAnalyzer* helps web developers by suggesting hints on the type of the different JS elements present in the page. This is done by providing an initial prediction on the JS functionality type, allowing the developer for an interactive enabling/disabling, and offering a live impact examination via a controlled browser. Using *JSAnalyzer*, the user can identify elements that have a visual impact on the page via the enabling/disabling feature. For example, if the JS element that is controlling a page menu is disabled, the menu would not function properly. However, there exist JS elements that have no visual effect on the page visual content or interactive functionality. These elements are commonly embedded in web pages for tracking and advertising purposes, and can fall in one of the known predefined JS categories [16]. These elements are tagged in the interface with the corresponding category (see example JS category tags in Figure 3). Categories are predicted with high accuracy using a supervised learning classifier, so that the user can have an insight into the purpose of an element that has no visual impact on the page content or functionality. The role of the classifier is not to directly disable a JS element, but to predict the script category and help the human developer to get an early insight into the script (in the form of suggestions/hints). The final decision is left to the user, who can display the JS code of a selected JS element via the interface to have a closer examination before making a decision.

We considered the ML model created and evaluated in [25] that uses the categories defined by the community of web experts, Web Almanac [16], which are: Advertising, Analytics, Social, Video, Utilities, Customer Success, Tag Management and Content. The model successfully predicts the JS elements with an f1-score of 88%, and a precision of 89%. While this may not be perfect, the aim of the JS classifier is merely to give hints to assist the user on type of the JS element. The user is the ultimate decider on what to remove or keep, in this sense JSAnalyzer is a user driven tool, and the user can always revert back the decision in case of a miss-labeled element. Additionally, we only consider labels which have a high confidence measure. If the confidence is low, we use a modified version of the rule-based classification presented in JSCleaner [26] to label these elements. This is done with one minor modification where the replaceable category is merged with the critical one. This is done to keep JSAnalyzer less invasive, not altering the JS elements code.

### 3.3 Dependency Tracking

JSAnalyzer tracks JS dependencies to group dependent elements together. Two types of dependencies are considered: functional dependency, and network dependency. A group of functionally dependent JS elements can be defined as elements that depend on each other to function properly, whereas a network dependent JS element is a JS resource that is being requested through another JS element. For example, the script that is embedded in another JS element can only be requested after that JS element is evaluated. JSAnalyzer utilizes the dependency tracking feature to ensure that JS dependencies are monitored, ensuring that cases where a certain JS element is enabled while one or more of its dependencies are disabled, or vise versa, do not arise. An element that seems to be non-critical might perform a request to fetch an element that is critical to the page. In this case, both elements should be considered critical and must be enabled to preserve the visual page content and functionality. Dependent JS elements are shown in groups (see example grouped elements in Figure 1) to avoid page breakage due to faulty disabling. Whenever the user enables an element, JSAnalyzer automatically enables the required dependent elements to preserve the functionality.

JSAnalyzer examines the relationship between different JS elements in a web page and draws a dependency graph to represent network dependencies among different JS elements in a web page. This is to achieve two goals:

- First, to ensure that cases where a certain JS element is enabled while one or more of its dependencies are disabled, or vise versa, do not arise. Dependent JS elements are shown in groups (see example grouped elements in Figure 3) to avoid page breakage due to faulty disabling. Whenever the user enables an element, *JSAnalyzer* automatically enables the required dependent elements to preserve functionality.
- Second, to consider re-classification when necessary. An element that seems to be non-critical might perform a request to fetch an element that is critical to the page. In this case, both elements should be considered critical and must be enabled to preserve the visual page content and functionality.

### 3.4 User Interface

As shown in figure 3, *JSAnalyzer* consists of an interactive user interface (see the right side of figure 3) that controls a web browser (see the left side of figure 3), via which the effect of user interactivity is visually reflected. The developer can access a page by requesting its corresponding URL via the interactive UI, and can request JSAnalyzer to provide an initial automatic analysis by clicking on "Analyze page" button (See Figure 3). JSAnalyzer waits until the page is fully loaded in the controlled browser window, by periodically investigating the changes in the Document Object Model (DOM). Once the page is fully loaded, the performance logs of network activities are parsed to extract the requested JS elements and provide an initial analysis before listing them to the user. The UI gives an overall listing of all JS elements embedded in a web page, and list the dependency of these elements on each other in a hierarchical manner. This tree-like representation gives the human analyzer a first look on all[1] the JS elements used in his page, as well as their inter-dependencies. The user can then interact with JS elements by enabling/disabling, and preview the effect of these interactions by clicking "Apply Selection". The browser then reloads a new version of the page according to what is enabled and disabled. Clicking the "Save and load simplified page" button allows for the generation of the lightweight reports, where enabled JS elements are preserved, while disabled elements are removed.

*JSAnalyzer* shows a label with the JS category prediction beside each JS elements in the UI. For the category prediction, each element embedded in the requested page is being evaluated by employed JS classifier (see Section 3.2). These labels are treated by the users as hints to aid the developer in their page optimization quest. Each JS element is represented to the user as a checkbox labeled with the name of the element, and tagged with a label as shown in Figure 3. Whenever a checkbox is clicked, the evaluator is given the chance to inspect the corresponding JS code, which is displayed in the source panel at the right side of the interface. A "checked" checkbox represents an enabled JS element, whereas an "unchecked" checkbox represents a JS element that is disabled. *JSAnalyer* initially "checks" the boxes of the elements that are predicted by as critical, whereas elements that are predicted as non-critical are "unchecked". *JSAnalyzer* considers the user preferences (identified in a text file) to determine critical and non-critical JS categories (for example, by default, Analytic is non-critical whereas Content is critical). The user is also given the option to select all JS elements at once for enabling or disabling via the "Select All" checkbox.

To produce a lighter version of a web page using *JSAnalyzer*, a user would experience the following steps:

**Pre-Analysis:** In this step, the default settings are altered to identify critical and non-critical categories according to the user preference. The default *JSAnalyzer* settings consider JS elements that are critical to the page visual content and/or interactive functionality as critical, and as non-critical otherwise. For example, analytic and Advertising are non-critical by default. The user enters the URL of a given page via the user interface (see 1 in Figure 3). *JSAnalyzer* responds by two actions: loading the requested page in the controlled browser (see 2 in Figure 3), and performing an initial JS analysis by invoking the employed classifier. The original page will also be loaded side by side (See Figure 1 A,B), so that the user can evaluate the simplified page with respect to the original version. Initial analysis results are visualized via the interface (see 3 in Figure 3). Critical elements are shown checked whereas non-critical elements are unchecked, and dependent elements are shown in groups.

**Interactive Analysis:** The user is able to interact with the page by disabling the enabled elements, one by one (see 4 in Figure 3). The "Apply Selection" button can be clicked after every interaction to reload the page with the user selections applied, in order to observe the impact of these selections on the page (see 5 in Figure 3). The user objective is to have the minimum number of enabled elements without sacrificing the visual page content or functionality. Thus, the user can re-enable the elements that results in missing page content or functionality when they are disabled. This would be an iterative approach, where the user is the ultimate decider on what to remove or keep.

**Mobile Page Generation and Evaluation:** When the user analysis is completed, *JSAnalyzer* provides the option to generate a simplified page based on the analysis, by clicking on the "Save and load simplified page" button (see 6 in Figure 3). In the simplified page, critical JS elements (with the checkboxes checked in the user interface) are preserved, while non-critical elements (with the checkboxes unchecked in the user interface) are removed. This way, *JSAnalyzer* provides a rapid generation of a lighter version from an existing page. This is achieved by reducing the number of JS elements embedded in the original page, without sacrificing the visual content or interactive functionality. The generated page is saved on the server and loaded to the user side by side with the original page to allow for qualitative evaluation.

---

[1] By all, we also refer to the JS elements that are being recursively requested, and that are not easily identifiable by the developer from the original index.html file.

*JSAnalyzer* employs two evaluation tools: PQual [33] that provides a similarity score of the newly created mobile page with respect to the original, and lighthouse [18] tool that is configured to provide quantitative evaluation reports, one for the original version (See Figure 1 D), and another one for the generated page (See Figure 1 E). These reports aim to provide an insight into the performance gain of the *JSAnalyzer page* in comparison to the original.

The output comes in the form of a modified HTML file, where non-critical inline scripts are removed from the HTML code, this is done for both inline JS code within a <script> tag, and for external JS files defined within the src attribute of a <script> tag. For the recursive JS files, we generate a report for the developer highlighting which JS files need to be blocked/removed, and where these scripts are being recursively called. The developer has the option to track these scripts and modify the code accordingly to block them.

### 3.5 Use Cases

There are two use cases for JSAnalyzer: if the developer is using a pre-designed template (generic pre-designed pages or a set of HTML pages that can be used by anyone) or a page builder (such as wordpress), it is a valid assumption that she/he won't be aware of the external JS code associated with their pages, and as a result using JSAnalyzer would benefit such developers. On the other hand, if we take the case of more experienced/professional web developers even though they have generally a better understanding of the used JS code, due to the excessive reliance on third party libraries, such as JQuery, there will be many additional recursive JS files that are being brought by these libraries due to the fact that these libraries are generic versions that is implemented to cover many cases and functionality. As a result, many additional JS files associated with that library would be brought in, even though the developer is not using that functionality of the JS library. This is different from conventional programming compilers, where these compilers are able to remove unnecessary code/functions that are not used by the main program. Unfortunately, the dynamic nature of JS prevents this from happening, because the JS code gets processed dynamically for the first time by the client browser.

## 4 IMPLEMENTATION

The user interface is implemented using wxPython [17] and Selenium WebDriver [14] along with the Chrome browser. The detection of external JS elements is performed at the browser level via Chrome's DevTools Protocol (CDP), which makes it possible to view all network activities and to block specific resources [9].

Network dependencies are tracked using Chrome DevTools APIs. The functional dependencies are tracked using Puppeteer [6], a Node library that is maintained by the Chrome DevTools, which allows for high-level control of Chrome and Chromium over the Chrome DevTools Protocol. The Chrome DevTools Protocol provides an array of tools to inspect, debug and profile Chrome and Chromium. One of these tools is the CPU Profiler, which aids JS dependecy tracking. Puppeteer can record the performance of a page and generate a CPU Profile. In the CPU Profile, the call stack of the JavaScript V8 engine is exposed and individual stack frames, as well as the relationships between them, can be examined. Each stack frame represents a function call and its parent stack frame.

For the proxy server implementation, we extended MITM Proxy [28] with a caching framework using its flexible scripting approach, where HTTP interception is controlled to interact with an SQL database server. The database is utilized to store the URLs of all HTTP requests, and the relative path of HTTP response files.

We use the extended MITM proxy implementation in three instances in our architecture. We developed a local MITM proxy at the client-side to allow for caching the requested URLs. The local caching facilitates the testing process by allowing the user to observe changes in real time. Moreover, it makes the testing process faster and cheaper as it avoids unnecessary internet requests for URLs that have been previously cached, as well as avoiding concurrency issues if multiple users are interacting with the tool.

The other two MITM proxy implementation are deployed in the remote server. The first one clones web pages in the aforementioned manner. Its main utility is to provide a stable data set that allows for reproducible results over time, as web pages can be periodically changed. We utilized this MITM proxy to clone a set of popular web pages [13] and make them available for analysis. The second MITM proxy implementation (operating on a different port) stores the list of blocked URLs (which consists of JS elements that are non-critical). This implementation allows for the visualization of the optimized version of a given web page, which is a key consideration for generating simplified pages and allowing for comparative performance evaluation.

## 5 EVALUATIONS

We performed three different types of evaluations of *JSAnalyzer*: a) a qualitative evaluation, b) a quantitative evaluation, and c) web developers' evaluation. The qualitative evaluation focuses on evaluating both the content and the functional similarity scores of the pages generated by *JSAnalyzer* in comparison to their original versions, while the quantitative evaluation focuses on assessing the standard page load time metrics along with the user experience timing metrics suggested by Google Lighthouse [18]. We compared against JSCleaner [26], which is a recent web complexity solution that generates simplified web pages from existing pages. In our evaluations, we refer to the following versions of a web page:

- *Original page*, which is the original web page that is accessible via mobile web.
- *JSCleaner page*, which is the page generated by JSCleaner.
- *JSAnalyzer page*, which is the page generated by JSAnalyzer.

The evaluation dataset consists of 100 popular web pages listed by Alexa [19], along with their 100 counterpart pages generated by *JSAnalyzer*, and by *JSCleaner*. These pages were made available via a proxy server for our analysis.

### 5.1 Qualitative Evaluation

Twenty two users were recruited from an international university campus by posting online ads on some of their popular social media groups. The recruited users were informed to spend a maximum of 30 minutes on evaluating the similarity of two pages that were shown side-by-side. The recruited users were not part of this work by any means beforehand. They all spoke English and came from different backgrounds and study disciplines. The user study was conducted online, with all the required explanations available on the evaluation page. Users were asked to evaluate as many pages as they could within 30 minutes. An institutional review board (IRB) approval was given to conduct the user study, and all the team members have completed the research ethics and compliance training (CITI [10] certified). The recruited users were asked to evaluate the similarity between the two versions of the pages (the original and the JSAnalyzer version), by answering the following questions (Figure 4a):

- Rate the content similarity (in a scale from 0 to 10)
- In the case of missing content, mention all types of missing contents that apply: (text, images, advertisements, video, layout/beautifiers, other embeds ( tweets, maps and etc.))

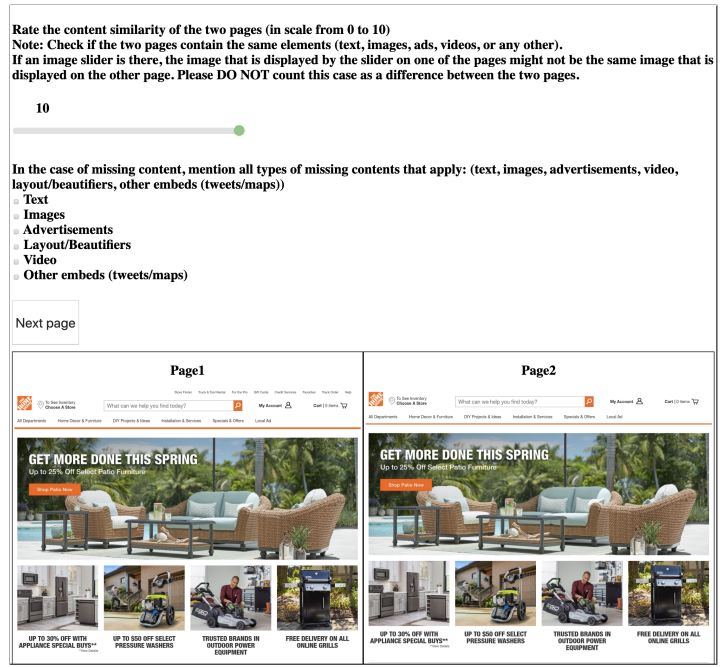

(a) Screenshot of the online user study

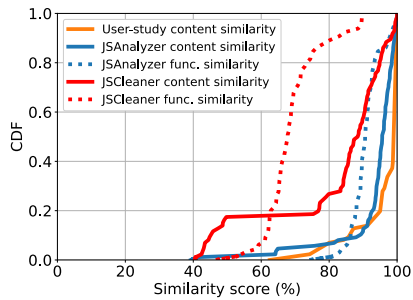

(b) PQual vs. user study similarity CDFs

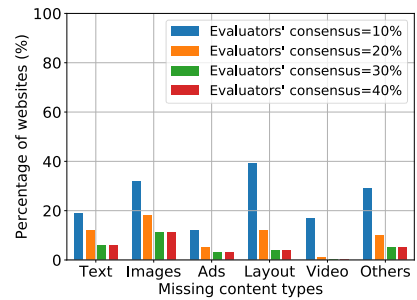

(c) User study: missing content percentages

Fig. 4. Qualitative evaluation results

Additionally, we computed both the content and functional similarity scores of the 100 *JSAnalyzer* pages and the *JSCleaner* pages with respect to the original versions using the *PQual* tool [33], which automates the qualitative evaluation of web pages using computer vision. Figure 4b shows the cumulative distribution functions (CDFs) of the content and the functional similarity scores for both *JSAnalyzer pages* and *JSCleaner pages* with respect to the original pages. Additionally, the figure also shows user-study content similarity scores for the *JSAnalyzer pages*.

The figure shows that the for almost 90% of the pages, *JSAnalyzer* achieves a content similarity score above 90%, which is proven by the user-study (orange curve) and the *PQual* content similarity scores (solid blue curve). Additionally, the results show that *JSAnalyzer* achieves functional similarity score of ≥ 90% for about 80% of the pages, whereas the rest of the 20% of the pages show a score of ≥ 80%, depicted by the dashed blue curve. In contrast, *JSCleaner* shows a significant qualitative degradation in both the content and functional similarity scores, depicted by the solid and dashed red curve, respectively. Specifically, *JSCleaner* achieves content similarity scores of ≥ 80% for 80% of the pages, however, for the rest of the 20% of the pages the content similarity scores are significantly low (between 40%-45%). Additionally, *JSCleaner* also degrades the functional similarity, achieving a score between 60% - 80% for almost 90% of the pages.

Figure 4c shows the results of the user study in terms of missing contents marked by users. The results show the percentage of web pages that had missing content per content type (text, image, video, layout, ads, and others). There are multiple bar charts that are shown here depending on the users' consensus. The consensus ratio is evaluated by computing the ratio of users that agreed on having a missing item in each page. For example, if only one user out of ten others (who have evaluated the same page) mentions that the page is missing an image, then the page is counted in the pages that are missing an image when the consensus is fixed to 10% (1 out of 10 people). However, when we increase the consensus ratio to 20% this page is not considered as missing an image. The results shows that even in the worst-case scenario with very low consensus score of 10% (depicted by the blue bar graph), most of the *JSAnalyzer* pages were not missing major content, apart from some missing layout and beautification elements (around 40%). However, as soon as we increase the user's consensus to 20% we can see a large drop in the number of pages that are missing items, even in the case of layout, which drops to about 10%. The consensus ratio can be viewed as a reflection of the user study confidence in the generated results.

### 5.2 Quantitative Evaluation

The quantitative evaluation consists of two main parts: a) the user experience using Google's lighthouse [18], and b) the standard page load time metrics using the webpagetest automation framework. Both evaluations were performed using real mobile phones: a low-end phone (Xiaomi Redmi Go), and a high-end phone (Samsung Galaxy S8+).

**a) User Experience Evaluation:** in this evaluation, we focused on three different metrics from the lighthouse report:

- First Meaningful Paint (FMP): which measures when the primary content of a page is visible to the user. It represents the time in seconds from when the user initiate the page load and the page rendering the primary above-the-fold content.
- Speed Index: which measures how quickly the page contents are visually displayed during page load.
- Time To Interactive (TTI): measures how long it takes a page to become fully interactive, either by displaying useful content, or by registering event handlers for the most visible elements.

Figure 5a shows the above lighthouse metrics for the 100 pages, using the Samsung high-end phone. The results are shown in the form of a Cumulative Distribution Function (CDF). In Figure 5a, the solid curves represent the results of the original pages, whereas the dashed ones show the results of the *JSAnalyzer* pages. The results show that there is a small gain in terms of the speed index (i.e., blue curve) with the *JSAnalyzer* pages, and almost a negligible gain in the FMP (i.e., red curve). However, the TTI of the *JSAnalyzer* pages (i.e, dashed green curve) shows a significant 40% gain

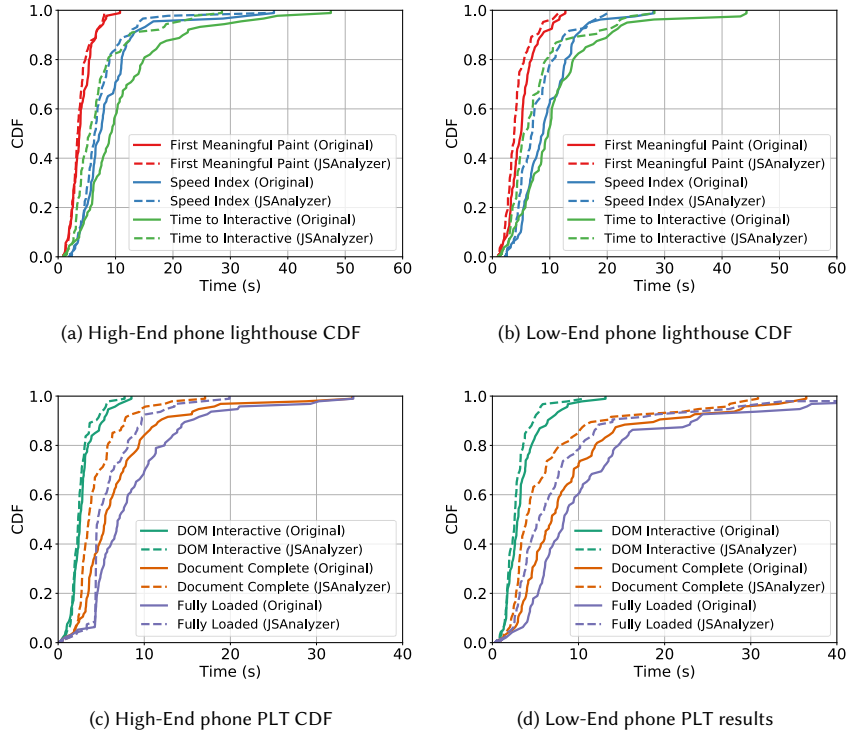

(a) High-End phone lighthouse CDF
(b) Low-End phone lighthouse CDF
(c) High-End phone PLT CDF
(d) Low-End phone PLT results

Fig. 5. JSAnalyzer Quantitative results

over the original pages (by observing the median value of the CDF). Figure 5b shows the low-end device's lighthouse metrics for the same 100 pages. Here, *JSAnalyzer* introduces a higher gain compared to the original page in almost all three metrics. This is evident by the gap between the different curves, i.e., the gap between the dashed and the solid curves. On median, *JSAnalyzer* achieves about 50% reduction in the time to interactive compared to the original pages.

Figure 6 shows the overall lighthouse performance score from 0 to 100 based on the real website performance data taken from the HTTP Archive. The results show the median value taken over all the 100 pages. A score of zero represents the lowest possible score, whereas a score of 100 indicates that the page is in the ninety-eighth percentile of websites for that metric. The figure shows that *JSAnalyzer* has managed to increase the performance score by more than 90% in the case of the low-end phone, and about 30% in the case of a high-end phone. In both cases, *JSAnalyzer* has actively enhanced the web pages score from a low value (represented by red) to an average performance (represented by orange). Although *JSCleaner* managed to increase the performance beyond *JSAnalyzer* in both low-end and high-end devices, it achieved this increase on the expense of losing similarity to the original page, as seen earlier in Figure 4b.

**b) Page Load Timings Evaluation:** similar to the user experience evaluation, we evaluated the different page load timings for our selected 100 popular pages using both: the low-end and the high-end phone devices. In this analysis, we focused on three main metrics: DOM interactive, Document complete, and Fully Loaded. DOM interactive marks the point when the browser has finished parsing all of the HTML and the DOM construction is complete. The difference between Document complete and Fully loaded is that Document complete represents the point when the browser onLoad event fires, which generally means that all of the page static content has loaded. While Fully loaded marks the point when the network has been idle for about 2 seconds, any activity beyond the Document complete comes from JavaScript loading some dynamic content. Figure 5c shows the CDF results for the high-end phone, whereas Figure 5d shows the low-end phone results. The high-end phone results tentatively shows a lower load times compared to the low-end one, due to the fact that the high-end phone possesses higher processing power and larger memory. However, in both cases it can be noticed that the *JSAnalyzer* pages achieves lower page load time compared to the original pages. In fact, the low-end phone has a reduction of more than 33% in the document complete and the fully loaded timings.

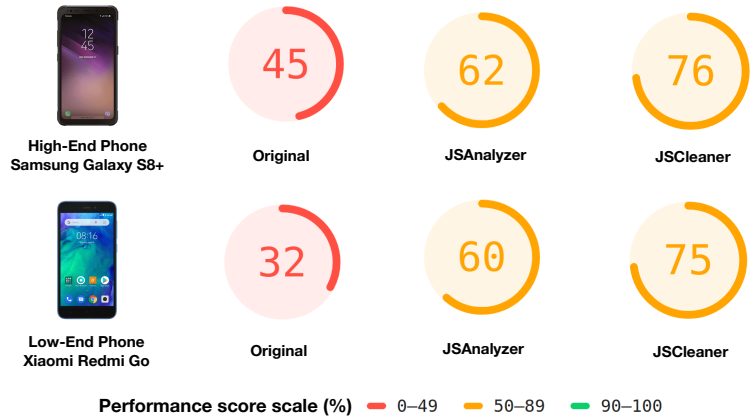

Fig. 6. Overall performance metric score

## 5.3 Comparison to Google AMP

In this subsection, we compare the JSAnalyzer pages to google AMP pages. Because AMP pages do not usually refer to their original counterpart pages, we were able to find only 21 AMP pages that have original counterpart pages. We started with the AMP pages listed in [37], which consists of many articles that belong to the same website. We filtered out these articles to select each website only once. We also filtered out the pages that do not have an original counterpart and ended up with a compiled list of 21 unique web pages. Although 21 pages might seem to be a small number, it does however gives a good insight on the differences between *JSAnalyzer* and Google AMP. To provide the necessary comparison, we analyzed each of the original pages in the list using *JSAnalyzer* to generate a simplified page. Then, we compared the performance of the original pages with their AMP and *JSAnalyzer* versions.

Figure 7 shows the evaluation results of the *JSAnalyzer* comparison to both Google AMP and the original pages. We evaluated four main quantitative metrics: First Contentful Paint, Speed Index, Size of resources (measured in MB), and number of network requests. The First Contentful Paint measures how long it takes the browser to render the first piece of DOM content after a user navigates to the page. This is shown in Figure 7a in the form of a CDF, where it can be seen that *JSAnalyzer*, represented by the orange curve, manages to render the first DOM content before both AMP (green curve) and the original page (blue curve). *JSAnalyzer* lowers down the Speed Index significantly compared to both AMP and the original pages. In fact, it can be seen that *JSAnalyzer* reduces the Speed Index for all pages below 12 seconds, compared to 32 seconds and 60 seconds for AMP and the original pages respectively.

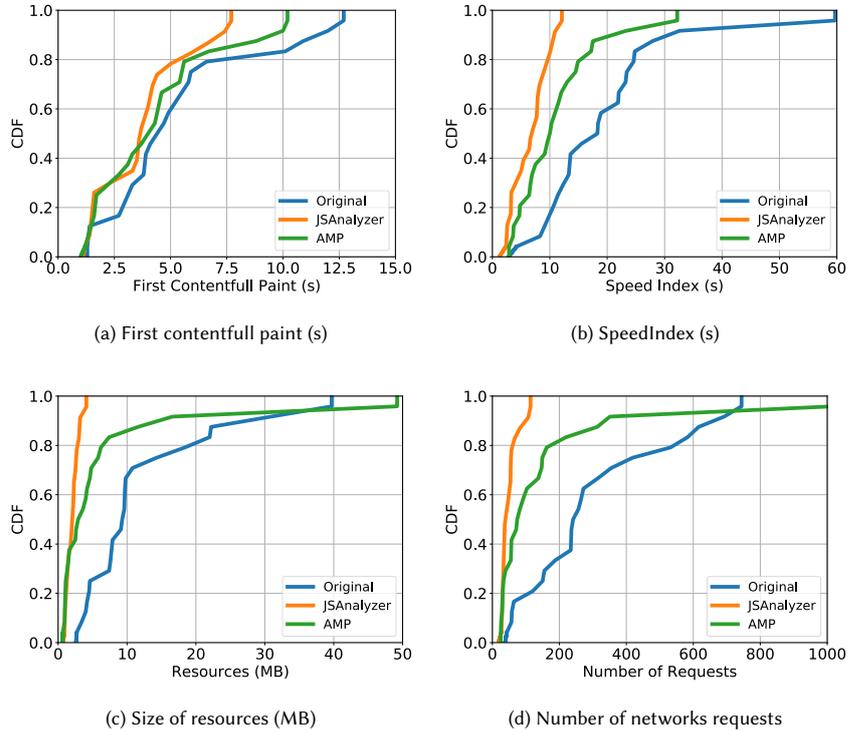

Fig. 7. JSAnalyzer comparison to AMP

Figure 7c and 7d show that *JSAnalyzer* lowers down the amount of resources and network requests drastically compared to both AMP and the original pages. In fact, for the upper 20 percentile of the CDF, *JSAnalyzer* reduces the size of resources between 45% to 92% in comparison to AMP. The same can be observed for the number of network requests. These are significant improvements over Google AMP, that help users not only save data cost, but also conserve energy by lowering down the amount of network requests as well as the transfer size of the data being brought over the network.

### 5.4 Web Developers' Evaluation

*JSAnalyer* was evaluated by 23 web developers at different levels (7 novices who can use existing templates for web development, 14 experienced who use templates regularly and can write limited JS code, and 2 expert developers working in web development and can write their own JS code from scratch). The developers were invited to evaluate the tool through social media groups. The developers were chosen based on their resume and web development portfolio, where we aimed to recruit developers with early to intermediate web development experience. Developers who accepted the invitation were asked to select an hour time-slot from a set of available times/dates for evaluation. During each evaluation session, the developer was introduced to the purpose of the tool and a had a quick tutorial on how to use it. Then, he/she was asked to select one or two pages for the evaluation. They were given access to the JSAnalyzer tool to analyze their pages, produce the simplified versions, and evaluate them. When the study was completed by all the developers, an anonymous survey form was sent by email to each of them asking to answer a set of questions. To avoid subjectivity, we informed the developers that the survey is anonymous and no respondent identification information is collected.

**User Evaluation Questions:** the developer survey consisted of three main parts: general questions, *JSAnalyzer* evaluation, and JavaScript questions. In the general section, developers were asked about the main purpose of their pages, and if they think that these web page can work well on mobile devices. In addition, they were asked if they have tried to evaluate their pages on mobile devices, and if they think that it is necessary to provide a lighter version of web pages for mobile users. *JSAnalyzer* questions included the following:

- Rate the tool, the usability of the tool, and its usefulness in a score from 0 to 10.
- Select all that apply, the tool helped me in the following: have a better understanding of JS elements in my web page, and the elements that might not be necessary for all users, create a lighter version of my web page, interact with JS elements in my page.
- Evaluate the full functionality of the generated page, and rate the functionality of the lighter version if not fully functional in a score from 0 to 10.
- Rate the similarity of the generated page with respect to the original version in a score from 0 to 10.
- Do you believe that the lighter version has a better performance compared to the original?
- Please provide a written feedback on your overall experience with *JSAnalyzer*.

**Developers' Evaluation Results:** Figure 8a shows the box plot results of the rating questions that we asked the web developers. The results show that the developers have rated the pages created by JSAnalzer in median scores of more than or equal to 80%. As for the 80% usability, a number of the developers suggested that the effect of enabling/disabling can be done a bit faster, in addition to a number of minor tool features improvements that can easily be implemented. The usefulness of the tool scored about 90%, and the majority of the developers agreed that the pages created by JSAnalyzer retains exactly the same content and functionality of the original pages.

Figure 8b shows the stack bar plot of a couple of the Yes/No questions that we asked the developers. Interestingly, the results show that 61% of the developers think that their web pages work well over mobile devices, however, only half of them have actually tested the pages on mobile devices. This is part of the issues that plagues mobile web pages, that many web developers simply assume that their pages would work on mobile devices, and that they don't need to make any changes in the original pages that were designed for desktops. Additionally, the results also show that 87% of the developers believe that it is necessary to provide lighter web pages for mobile users, however only 21% of them are aware of some solutions that can help in generating lighter versions of their pages. In this regard, many of the developers have found that *JSAnalyzer* is a very useful and unique tool that can help in analyzing and simplifying their pages, thus creating a lighter versions for mobile devices, given that extra libraries are sometimes being added without the developer's explicit consent.

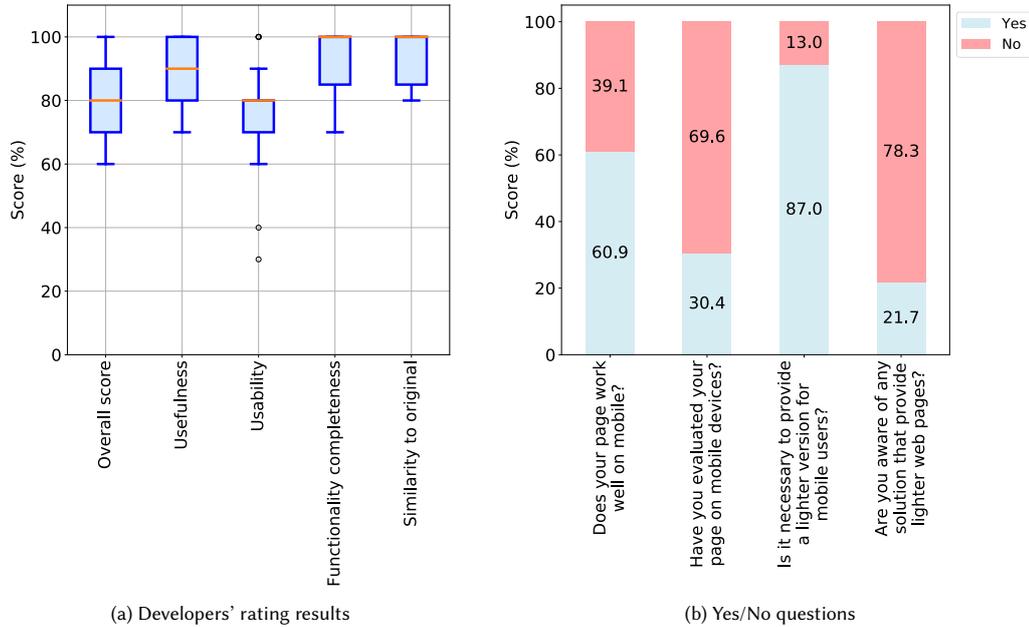

(a) Developers' rating results  (b) Yes/No questions

Fig. 8. JSAnalyzer evaluation by web developers

## 6 DISCUSSIONS

### 6.1 JSAnalyzer Limitations

With the current implementation of *JSAnalyzer*, the utilization of the *PQual* evaluation tool is performed only at the end of an analysis session, in order to generate the qualitative scores for both the content and functional similarity of the optimized page in comparison to the original. Consequently, the user is unable to assess the quality of the intermediary created pages while disabling/enabling JS elements in the original page. An extended version of *JSAnalyzer* can employ *PQual* to assess the impact of each user interactivity, in order to assist the developer in recognizing the effect of disabling a given JS element. This can be performed by taking two screenshots of the page (before and after disabling the element), and then provide a quality score while highlighting the impacted area in the page.

Another current limitation of *JSAnalyzer* is that the tool does not aid the developer in discovering the actual location or functionality that got affected based on his/her interactions with the tool. Instead, the tool relies on the developer to visually inspect the affected area, or to manually test the missing functionality. An extension of the tool can be made to visually assist the developer in recognizing the exact location of the affected area, by displaying visual clues on the web page at these locations, thus enhancing the overall user experience of the tool.

### 6.2 JSAnalyzer Usability and Benefits

We outline some of the key qualitative observations made by the developers after they finished the developer evaluation study. One of the main objectives of *JSAnalyzer* is to provide an easy-to-use tool for web developers to create simplified versions of existing pages. Several developers shared highly positive thoughts on the usability of the tool as best summarized in one of the user quotations below:

"It is a very easy to use tool. I did not face any difficulty getting used to the format of the tool and I think that it was user friendly. Even if a person does not have a lot of knowledge about JavaScript, he can get used to using it pretty quickly. I also found the analysis part very helpful in gauging the effectiveness of the tool.".

The easiness of the tool is a key feature that aids the rapid generation of simplified pages for mobile users. Several developers also found that *JSAnalyzer* provides insights to the developer into each JS element in a web page by means of the category prediction aspect of the tool. This is summarized in one such quote below:

"I think the tool was incredibly easy to use and was relatively intuitive and easy to pick up. The labels showcasing what category each script belongs to and the tree-like layout of dependencies made it incredibly easy to fine tune the experience of the simplified site."

Several developers acknowledged the main benefits of *JSAnalyzer* page in improving the PLT of new pages and the ease of developing new versions of existing pages:

"The tool has a huge benefit in improving the load times of web pages"

"I think JSAnalyzer is an amazingly helpful tool for better web development."

The new simplified versions created by web developers can improve the browsing experience, especially for users with low-end settings:

"I feel like this tool definitely has lots of potential in that it isolates and removes unnecessary JavaScript elements. From the results that were displayed at the end, there is a noticeable increase in performance speed and other details. If this is able to work on the majority of pages, this would definitely help users with low-end settings."

Another key benefit of *JSAnalyzer* is the ability to filter out JS elements that might be added without the developer intention, for example, when the software development tool adds third-party libraries by default.

"Such a tool is nowadays really important. Usually these types of analysis are done on static code before the website is deployed, but most of the time extra functionalities (and extra libraries) are added without website creator's explicit consent. This is particularly true in Wordpress websites where installed plugins might bring unwanted library, or they might re-include libraries unnecessarily. JSAnalyzer seems really good to solve these issues, despite the fact that more features might bring more power and accuracy (e.g. in the classification part) to the analysis that are currently done."

### 6.3 Further Improvement

One of the developers has suggested: "I appreciate the simplicity in JSAnalyzer's interaction and reaction. Although the purpose of this tool is to create a lighter version of the web page, I noticed that the indication of copyright/rights of the page is also taken into analysis. I want to agree with reducing indented scripts, but copyright scripts should be seen as part of essential requirements that maybe should be red-flagged to the users.". Another suggested improvement was to have a web version of the tool: "Would like to see a web version of the tool.", "I think this tool can also be developed as an extension (like a chrome extension) so that it would automatically optimize a web page.". Finally, some developers were interested in having the visual impact automatically highlighted by *JSAnalyzer*: "The visual dissimilarities should be automatically detected by the tool."

## 7 CONCLUSIONS

*JSAnalyzer* is an analysis environment that aims to optimize JavaScript usage in modern web pages. This analysis can aid web developers and content providers in making optimization decisions on their pages to create simplified versions for mobile web. In this paper, we demonstrated how the optimized *JSAnalyzer* pages outperforms the original pages in multiple aspects, while maintaining the visual content and the interactive functionality of these pages.